# IEEE Big Data Cup 2022: Privacy Preserving Matching of Encrypted Images with Deep Learning

Vrizlynn L. L. Thing
Singapore
vriz@ieee.org

*Abstract*—Smart sensors, devices and systems deployed in smart cities have brought improved physical protections to their citizens. Enhanced crime prevention, and fire and life safety protection are achieved through these technologies that perform motion detection, threat and actors profiling, and real-time alerts. However, an important requirement in these increasingly prevalent deployments is the preservation of privacy and enforcement of protection of personal identifiable information. Thus, strong encryption and anonymization techniques should be applied to the collected data. In this IEEE Big Data Cup 2022 challenge, different masking, encoding and homomorphic encryption techniques were applied to the images to protect the privacy of their contents. Participants are required to develop detection solutions to perform privacy preserving matching of these images. In this paper, we describe our solution which is based on state-of-the-art deep convolutional neural networks and various data augmentation techniques. Our solution achieved 1$^{st}$ place at the IEEE Big Data Cup 2022: Privacy Preserving Matching of Encrypted Images Challenge.

*Keywords* — *privacy preservation, masking, encoding, homomorphic encryption, deep learning, convolutional neural networks*

## I. INTRODUCTION

The increasing adoption of technologies has transformed the world into a digitalized ecosystem, and the scale of digitalization has been expanding constantly and exponentially. In smart cities, we are seeing increasing adoption and deployment of smart sensors, devices and systems. These deployments have brought about improved physical security and safety for citizens through enhanced crime prevention, and fire and life safety protection. However, to do so, data that contain personal identifiable, sensitive and thus, potentially privacy invading information, would need to be collected.

There are many existing privacy laws and regulations implemented in different countries to instill privacy conformation. The General Data Protection Regulation (GDPR) [1] is a regulation that was introduced in 2016 and implemented to become enforceable in 2018, for data and privacy protection in member states in the European Union. The framework covers key data protection and privacy principles such as the rights of the data subject, data collection and processing matters, and the penalties for violating the regulation. Additionally, GDPR is applicable to non-member states if they are doing business in the EU and processing data of EU citizens. As such, the GDPR is widely referenced by other countries. Subsequently, laws and regulations pertaining to privacy were also introduced in other countries and states, such as Brazil's *Lei Geral de Protecao de Dados (LGPD)* [2]*,* the *California Consumer Privacy Act (CCPA)* in California State of the United States [3], the *Personal Information Protection Law (PIPL)* in China [4], and the *Personal Data Protection Act (PDPA)* in Singapore [5].

Thus, there is increasing awareness, plans and actions taken by organizations to ensure the privacy protection of consumers and citizens in general. One important technological approach to support these initiatives would be the application of strong encryption and anonymization techniques to the collected data, to preserve the privacy and enforce the protection of personal identifiable information.

In the IEEE Big Data Cup 2022 Challenge on "Privacy Preserving Matching of Encrypted Images" [6], different masking, encoding and homomorphic encryption techniques were applied to the images to protect the privacy of their contents. Challenge participants develop detectors to carry out privacy preserving matching of pairs of given images. In this challenge, we design and develop detectors based on various data augmentation and deep convolutional neural networks (CNN). Our proposed work achieves 1$^{st}$ place at the challenge.

This paper is structured as follow. In Section II, we introduce the convolutional neural network (CNN) architectures used in this work. In Section III, we introduce the challenge, the dataset and in particular, the privacy preserving techniques applied to the images. In Section IV, we describe our proposed approach and developed techniques. In Section V, we present the evaluation results and our findings. We conclude the paper in Section VI.

## II. CONVOLUTIONAL NEURAL NETWORKS

In recent years, CNNs have been demonstrated to exhibit outstanding performance through its powerful learning ability in computer vision, image processing benchmarking competitions, and natural language processing tasks [7]. CNNs leverage on multiple feature extraction stages to automatically learn data representations and have a strong ability to capture signal spatiotemporal dependences. Recent advancements focus on research on different activation and loss functions, parameter optimization, regularizations and most importantly, in CNN's architectural innovations. Significant improvements in the representational capacity of deep CNNs were demonstrated through architectural innovations.

LeNet was the first CNN architecture and was introduced in 1989 [8]. It was also being referred to as LeNet-5 and was a simple CNN which applied back propagation to handwritten zip code recognition. Since then, other CNN architectures such as AlexNet [9] which won the classification and localization tasks at the Large Scale Visual Recognition Challenge 2012 [10] with a deeper CNN model and more channel consideration, InceptionNet [11] that incorporated multi-scale feature extraction and increased the model width with varying sizes of kernels in parallel (instead of only depth increase), VGG (Visual Geometry Group at University of Oxford) [12] which used an architecture with very small (3x3)

*Figure 1: Example Network Architectures. Left: VGG-19, Middle: plain network with 34 parameter layers, Right: residual network with 34 parameter layers*

convolution filters and pushed the depth to 16-19 weight layers showed significant improvements over prior works in 2014, had also been proposed.

In 2016, ResNet [13] was proposed to address the vanishing gradient problem with deeply stacked multi-layers CNNs. Residual connections were introduced to create alternate paths for the gradient to skip the middle layers and reach the initial layers, which allowed extremely deep models with good performance to be trained. A ResNet architecture with 34 parameter layers, with comparison examples from VGG-19 and a plain network, from [13] is shown in Figure 1.

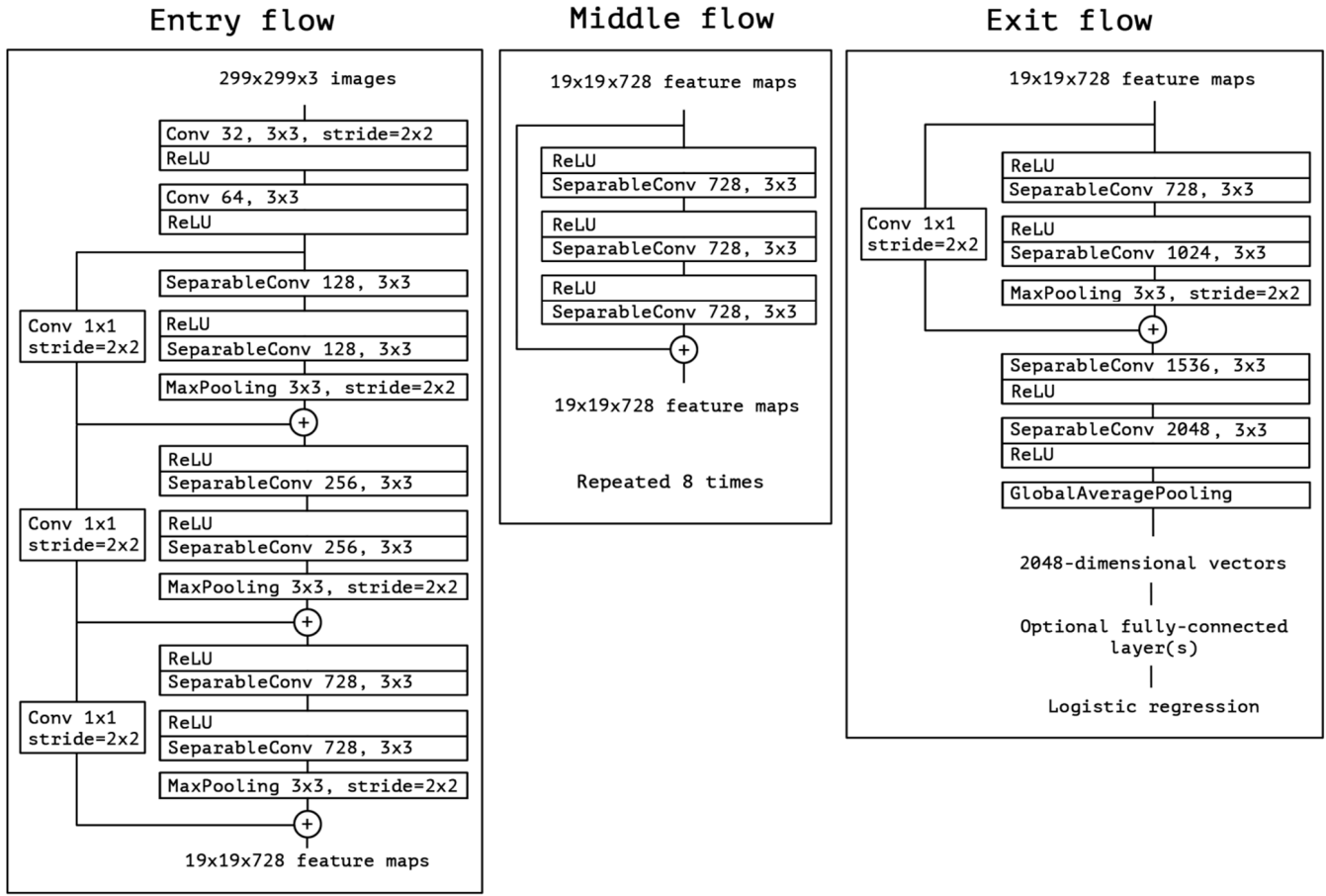

*Figure 2: XceptionNet Architecture: data goes through entry flow, then the middle flow which is repeated eight times, then through the exit flow. All SeparableConvolution layers use a depth multiplier of 1 (no depth expansion)*

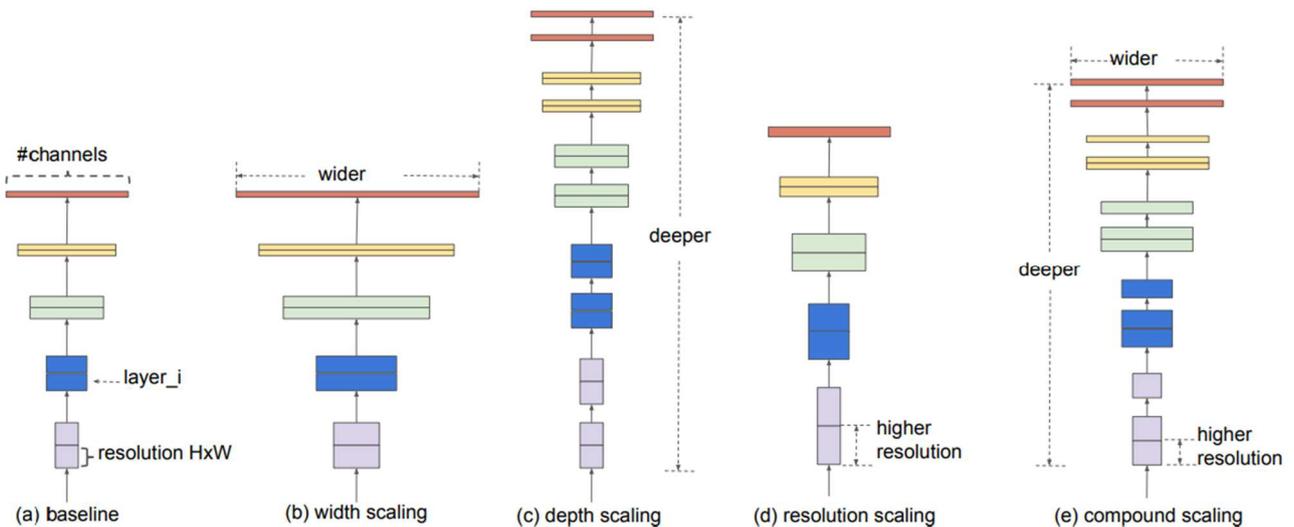

*Figure 3: EfficientNet Model Scaling. (a) baseline network example, (b-d) conventional scaling that increases one dimension of network width, depth, or resolution, (e) compound scaling method that uniformly scales all three dimensions with fixed ratio*

In 2017, XceptionNet [14], inspired by InceptionNet, was proposed. In XceptionNet, the Inception modules were replaced by depthwise separable convolutions, and an evaluation on ImageNet showed that XceptionNet was able to achieve better performance over InceptionNet-V3. The XceptionNet architecture from [14] is shown in Figure 2.

In 2019, EfficientNet [15] based on the design of balancing the network depth, width and resolution, and a method to uniformly scale all dimensions of depth, width and resolution using a compound coefficient was proposed. The compound scaling approach from [15] is shown in Figure 3. A family of eight model architectures, EfficientNet B0 to EfficientNet B7, scaled from the EfficientNet B0 baseline network were created. The authors demonstrated that EfficientNet B7 achieved the highest top-1 accuracy of 84.3% compared to InceptionNet-V4 at 80.0%, XceptionNet at 79%, ResNet152 at 77.8%, and ResNet50 at 76%, on the ImageNet data.

In this work, we will leverage the ResNet (specifically ResNet152, that is, ResNet with 152 parameter layers), XceptionNet and EfficientNet architectures to train our detection models.

### III. PRIVACY PRESERVING MATCHING OF ENCRYPTED IMAGES CHALLENGE

The scope of this challenge is related to the construction of next generation smart monitoring systems. These systems are being developed by MyLED [16] and QED software [17] as part of the AraHUB initiative [18]. However, an important consideration in this work is the preservation of privacy and protection of personal information (e.g. name, gender, age, location) of the persons in the captured data. Therefore, in this challenge, the aim is set to check if the masking, encoding and homomorphic encryption techniques being employed are sufficiently secure or robust.

Participants are provided a training data set consisting of original images and their encoded version. On a separate set of pairs (i.e. test data set), participants have to detect whether the encoded file contains the visible image for each pair. The challenge comprises of three subtasks, differentiated by the masking, encoding and homomorphic encryption methods used.

In the first subtask, the encoded image is the result of applying an image obfuscation scheme based on the chaotic system theory. The original image is cropped, padded, scaled to a square of 512x512 pixels, and divided into 32x32-pixel tiles. The tiles are then individually encoded using the Arnold's cat map [19] and formed as the final encoded images. An example of the original and its corresponding encoded image from the first subtask is shown in Figure 4.

In the second subtask, the same image obfuscation technique based on the chaotic system theory in the first subtask, is applied on the entire 512x512-pixel original images. Thus, the encoded images are visibly harder for the human eyes to identify with the original ones. An example of the original and its corresponding encoded image from the second subtask is shown in Figure 5.

In the third subtask, the original images are cropped to a square that contains a single human face only. The images are then scaled to 52x52 pixels in size. Image obfuscation based on the Brakerski-Fan-Vercauteren (BFV) Homomorphic Encryption (HE) scheme [20] is then applied to the images to generate the encrypted images (in the form of binary files). The BFV HE scheme allows addition and multiplication computations on the encrypted data directly.

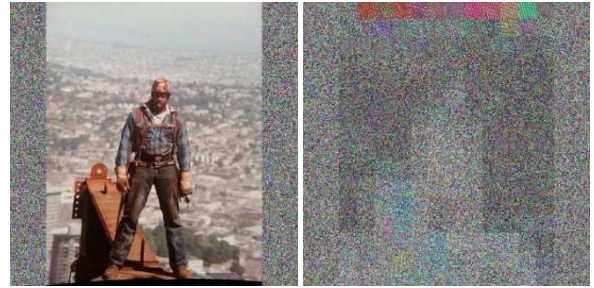

*Figure 4: Example from First Subtask*

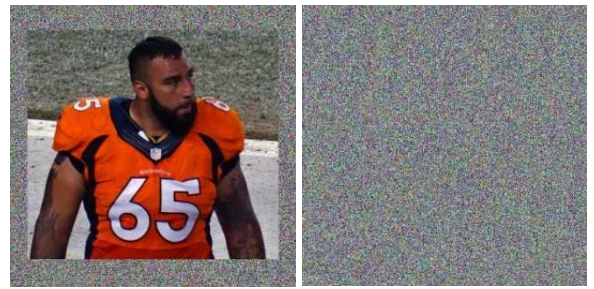

*Figure 5: Example from Second Subtask*

For each subtask, there is a training set comprising 10000 matching original-encoded pairs. For each subtask, there is also a test set comprising either matching or non-matching 10000 original-encoded pairs. Participants are required to create detectors using the training set and then analyze each of the 30000 test image pairs to perform detection. Participants then submit a text file containing 30000 lines (i.e. one for each test pair) to indicate if the prediction value for each pair is either 0 (not the same) or 1 (the same, i.e. a match).

### IV. PROPOSED DETECTION APPROACH

We firstly need to process the training data as the provided datasets contain only matching original-encoded pairs. For each subtask, we create another 10000 non-matching pairs by randomly choosing an encoded image for each original image to form non-matching original-encoded pairs. We then split the provided 10000 matching pairs and the generated 10000 non-matching pairs by 80:20 ratio, to create the model training dataset and validation dataset. We do the same for the other two subtasks.

For each training sample in the first and second subtasks, we form three channels from the original image and three channels from the encoded image. During training, we perform data augmentation randomly. The data augmentation methods include "shift-scale-rotate", "horizontal-flip", "random-brightness-contrast", "motion-blur", "gauss-noise", "to-gray", and "image-compression". We show examples of the data augmentation on the training dataset in Figure 6. We then perform normalization to each image, stack each pair of the three-channel images into six-channel images. and train the CNN models.

For the first subtask, the CNN architectures we consider are ResNet152 [13], EfficientNet B7 [15] and XceptionNet [14]. For the second subtask, the CNN architectures we consider are XceptionNet and EfficientNet B0.

For each training sample in the third subtask, we form three channels from the original image. As the corresponding encoded file is not an image but a binary file, we firstly convert the corresponding binary encoded file into a three-channel image before the subsequent processing. In this third subtask, we apply three different approaches to the model training.

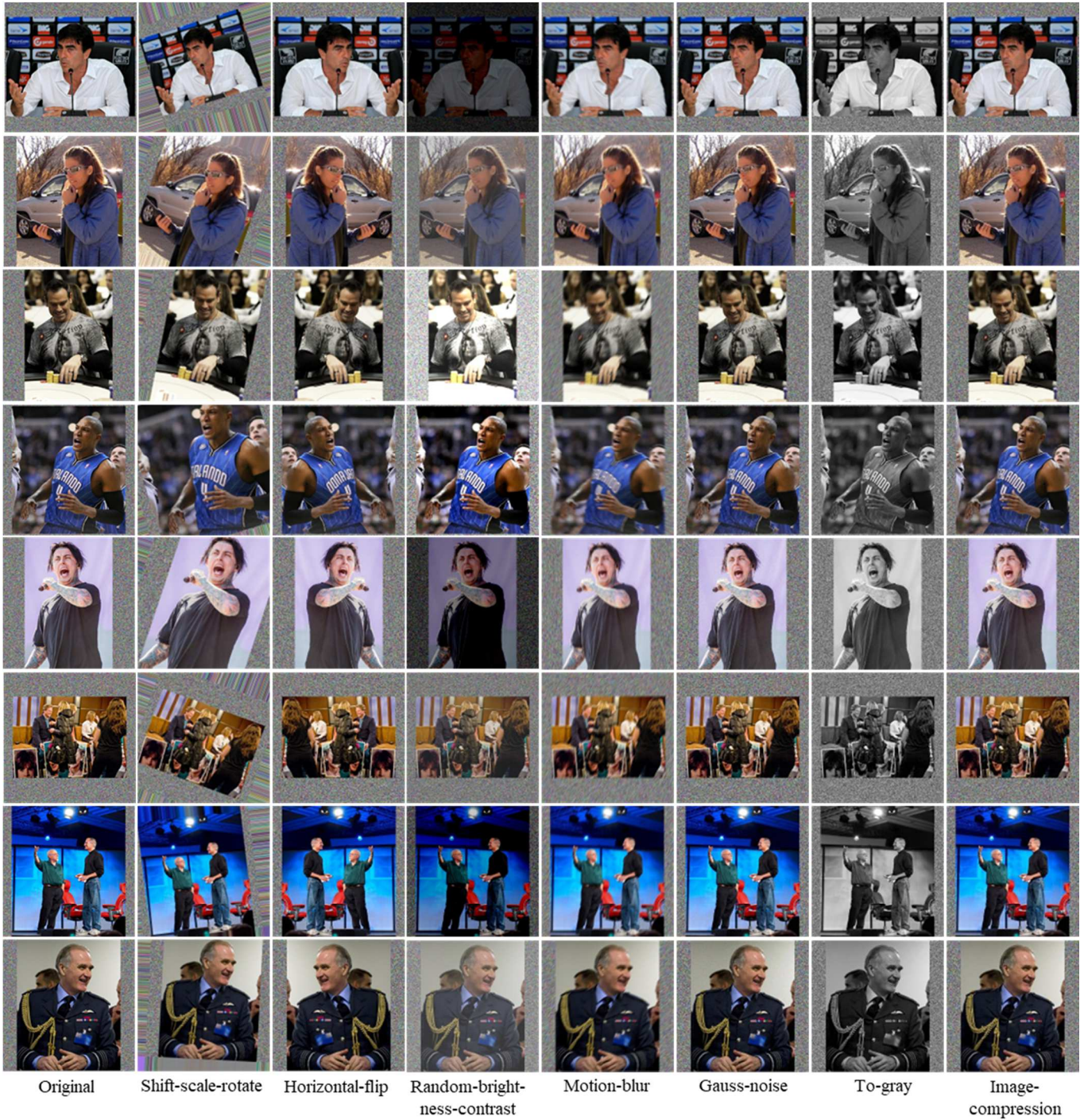

*Figure 6: Data Augmentation for First and Second Subtasks*

In the first approach, we perform data augmentation on the original image randomly at probability of each data augmentation method of 0.1. The data augmentation methods include "shift-scale-rotate", "horizontal-flip", "random-brightness-contrast", "motion-blur", "gauss-noise", "to-gray", and "image-compression". We show examples of the data augmentation in Figure 6. We then perform normalization to the original image. For the corresponding encoded image, we simply perform a division of the pixel values by 255. We then stack each pair of the three-channel images into six-channel images. and train the CNN models.

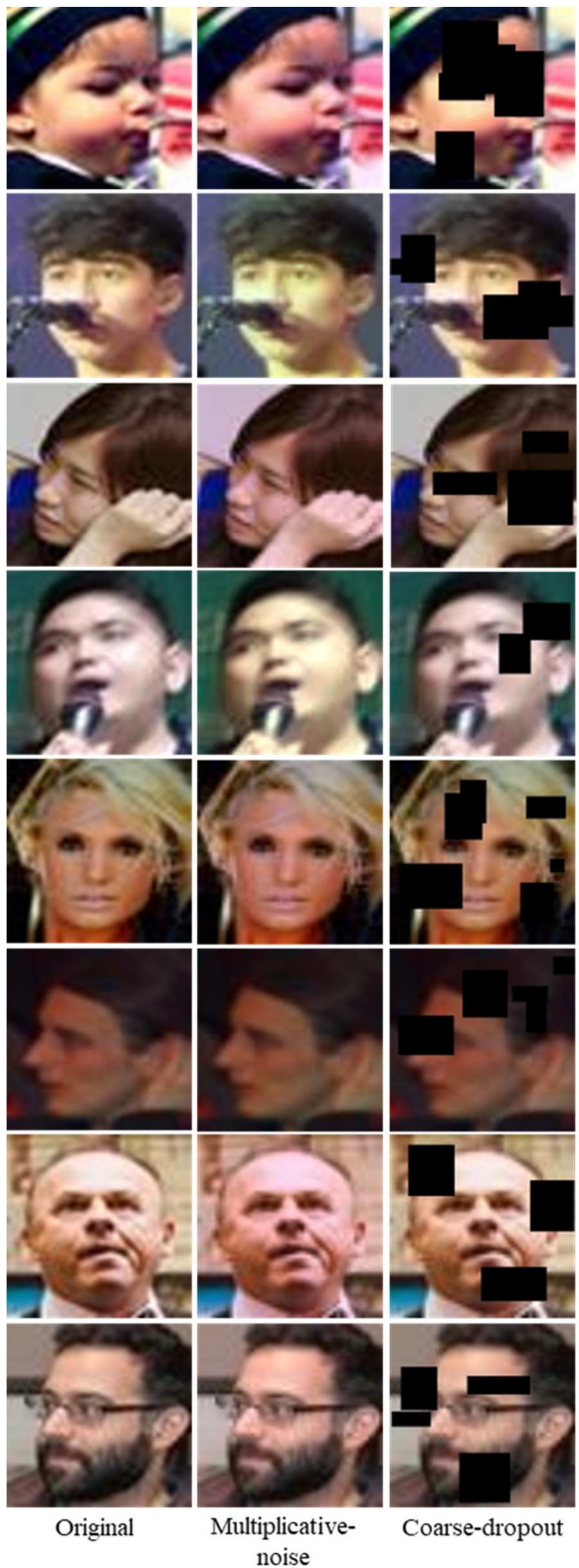

Figure 7: Additional Data Augmentation for Third Subtask

In the second approach, we simply perform a division of the pixel values for all the images by 255. We then stack each pair of the three-channel images into six-channel images. and train the CNN models.

In the third approach, we perform additional data augmentations (compared to the first approach) and on both the original and encoded images randomly. The probability of apply each data augmentation method to each image is at 0.1. The data augmentation methods include "shift-scale-rotate", "horizontal-flip", "random-brightness-contrast", "motion-blur", "gauss-noise", "to-gray", "image-compression", "multiplicative-noise" and "coarse-dropout". We show examples of the additional "multiplicative-noise" and "coarse-dropout" data augmentation on the third subtask's training dataset in Figure 7. We then perform normalization to the images, stack each pair of the three-channel images into six-channel images. and train the CNN models.

For the third subtask, the CNN architectures we consider are XceptionNet and EfficientNet B3 for the first training approach, and XceptionNet for the second and third training approaches.

We train all layers of the architectures from scratch. For subtasks where we train multiple models from the same architecture, we split the training and validation dataset using different random seed values. Models showing top proximate validation results are chosen to be included for the challenge submissions.

In the following section, we present the evaluation results and discuss our findings.

## V. RESULTS AND DSICUSSIONS

In Table 1, we show the models created for each subtask and their corresponding validation accuracy and loss. For the third subtask, we indicate 3-T1, 3-T2, 3-T3, for the first, second and third training approach, respectively. We create 4 models, 6 models and 5 models for the first, second and third subtask, respectively. We can see that each subtask has increased difficulty compared to the previous.

In this challenge, the evaluation by the organizer is carried out as a weighted piece-wise accuracy measure. The results on the leaderboard are computed as:

$$Acc = w_1 \times Acc_1 + w_2 \times Acc_2 + w_3 \times Acc_3$$

, where ($Acc_1$, $Acc_1$, $Acc_3$) and ($w_1$, $w_2$, $w_3$) are the accuracy (proportion of correct predictions) and weights for the first, second and third subtask, respectively. The rationale is that the organizer would like to encourage focus on solutions that work well on harder subtasks. Thus, the weights which are $w_1$, $w_2$, and $w_3$ are adjusted such that $w_1+w_2+w_3=1$ and $Acc \in [0,1]$. The weight values are assigned as follow.

For first subtask: $w_1 = 0.1$

For second subtask: $w_2 = 0.3$

For third subtask: $w_3 = 0.6$

The preliminary evaluation for the leaderboard is carried out on a small fraction of the test cases. For the final evaluation, the accuracy is calculated on the remaining part of the test data set.

In our solution, we compute the probabilistic scores of each pair from each of our created models for each subtask, and the scores were averaged and rounded to form the prediction result for each test pair. With our solution, we achieve a preliminary score of 0.7204 and a final score of 0.6946 on the leaderboard. The challenge organizer has published a summary report [21] on this challenge.

We observe that the first and second subtasks are relatively less challenging than the third subtask. In fact, based on the results among all participants in this challenge where all the final scores are below 0.7, the third subtask appears to remain an unresolved challenge. Further work and exploration to this subtask is warranted.

| Subtask | CNN Architecture | Model ID | Validation Accuracy | Validation Loss |
|---|---|---|---|---|
| 1 | ResNet152 | 1-1 | 0.992009 | 0.0233274 |
| 1 | EfficientNet B7 | 1-2 | 0.994722 | 0.017557 |
| 1 | XceptionNet | 1-3 | 0.998234 | 0.006196 |
| 1 | ResNet152 | 1-4 | 0.999 | 0.003773 |
| 2 | XceptionNet | 2-1 | 0.943464 | 0.172519 |
| 2 | EfficientNet B0 | 2-2 | 0.947596 | 0.158589 |
| 2 | EfficientNet B0 | 2-3 | 0.950458 | 0.146256 |
| 2 | XceptionNet | 2-4 | 0.955643 | 0.14449 |
| 2 | XceptionNet | 2-5 | 0.959497 | 0.131707 |
| 2 | XceptionNet | 2-6 | 0.954363 | 0.136174 |
| 3-T1 | EfficientNet B3 | 3-1 | 0.515491 | 0.854201 |
| 3-T1 | XceptionNet | 3-2 | 0.520897 | 0.69289 |
| 3-T1 | XceptionNet | 3-3 | 0.524585 | 0.69336 |
| 3-T2 | XceptionNet | 3-4 | 0.516292 | 0.693152 |
| 3-T3 | XceptionNet | 3-5 | 0.515097 | 0.693276 |

*Table 1: Detection Models and Validation Performance*

## VI. CONCLUSIONS

In this work, we explored the challenge of performing privacy preserving matching of encrypted images. The images were processed with different masking, encoding and homomorphic encryption techniques to protect the privacy of the image contents. We designed and developed detection models to carry out privacy preserving matching of pairs of these original-encoded images. We considered various data augmentation methods such as "shift-scale-rotate", "horizontal-flip", "random-brightness-contrast", "motion-blur", "gauss-noise", "to-gray", "image-compression", "multiplicative-noise" and "coarse-region-dropout" to generate additional data due to the limited dataset, and also leveraged on state-of-the-art deep convolutional neural networks such as ResNet152, XceptionNet and EfficientNet. Our proposed work won 1st place at the challenge at a leaderboard final score of 0.6946.

## ACKNOWLEDGMENT

The author would like to thank Zed Z. Dai, Randolph C. Loh and Jonathan W. Z. Lim for their hardware infrastructure support.

## REFERENCES

[1] "General Data Protection Regulation," [Online]. Available: https://gdpr.eu/.

[2] "Brazil's Lei Geral de Proteção de Dados," [Online]. Available: http://www.planalto.gov.br/ccivil_03/_Ato2015-2018/2018/Lei/L13709.htm.

[3] "California Consumer Privacy Act," [Online]. Available: https://oag.ca.gov/privacy/ccpa.

[4] "China's Personal Information Protection Law," [Online]. Available: http://www.npc.gov.cn/npc/c30834/202108/a8c4e3672c74491a80b53a172bb753fe.shtml.

[5] "Singapore's Personal Data Protection Act," [Online]. Available: https://www.pdpc.gov.sg/Overview-of-PDPA/The-Legislation/Personal-Data-Protection-Act.

[6] "IEEE Big Data Cup 2022: Privacy Preserving Matching of Encrypted Images Challenge," [Online]. Available: https://knowledgepit.ml/privacy-preserving-matching-of-images/.

[7] Z. Li, F. Liu, W. Yang, S. Peng and J. Zhou, "A survey of convolutional neural networks: analysis, applications, and prospects," *IEEE Transactions on Neural Networks and Learning Systems*, 2021.

[8] Y. LeCun, B. Boser, J. S. Denker, D. Henderson, R. E. Howard, W. Hubbard and L. D. Jackel, "Backpropagation applied to handwritten zip code recognition," *Neural computation*, vol. 1, no. 4, pp. 541--551, 1989.

[9] A. Krizhevsky, I. Sutskever and G. E. Hinton, "Imagenet classification with deep convolutional neural networks," *Advances in neural information processing systems*, vol. 25, 2012.

[10] "Large Scale Visual Recognition Challenge 2012," Online, accessed April 2022. [Online]. Available: https://image-net.org/challenges/LSVRC/2012/results.html.

[11] C. Szegedy, W. Liu, Y. Jia, P. Sermanet, S. Reed, D. Anguelov, D. Erhan, V. Vanhoucke and A. Rabinovich, "Going deeper with convolutions," in *IEEE conference on computer vision and pattern recognition*, 2015.

[12] K. Simonyan and A. Zisserman, "Very deep convolutional networks for large-scale image recognition," *arXiv preprint arXiv:1409.1556*, 2014.

[13] K. He, X. Zhang, S. Ren and J. Sun, "Deep residual learning for image recognition," in *Proceedings of the IEEE conference on computer vision and pattern recognition*, 2016.

[14] F. Chollet, "Xception: Deep learning with depthwise separable convolutions," in *Proceedings of the IEEE conference on computer vision and pattern recognition*, 2017.

[15] M. Tan and Q. Le, "Efficientnet: Rethinking model scaling for convolutional neural networks," in *International Conference on Machine Learning*, 2019.

[16] "MyLED," [Online]. Available: https://myled.pl/.

[17] "QED Software," [Online]. Available: https://qed.pl/.

[18] "AraHUB," [Online]. Available: https://arahub.ai/.

[19] G. Zhi-Hong, H. Fangjun and G. Wenjie, "Chaos-based image encryption algorithm," *Physics Letters A,* vol. 346, no. 1-3, pp. 153-157, 2005.

[20] J. Fan and F. Vercauteren, "Somewhat Practical Fully Homomorphic Encryption," *Cryptology ePrint Archive: Paper 2012/144, https://eprint.iacr.org/2012/144,* 2012.

[21] A. Janusz, M. Szczuka, B. Cyganek, J. Grabek, L. Przebinda, A. Zalewska, A. Buka and D. Slezak, "IEEE Big Data Cup 2022 Report: Privacy Preserving Matching of Encrypted Images," in *IEEE International Conference on Big Data, BigData 2022, Osaka, Japan, December 17-20, 2022*, 2022.